# Light confinement in stratum corneum

Gennadi Saiko

Department of Physics, Toronto Metropolitan University, Toronto, Canada

gsaiko@torontomu.ca

Abstract

The epidermis is the outermost layer of the skin, and it plays a crucial role in protecting the body from external insults such as UV radiation and physical trauma. The stratum corneum is the topmost layer of the epidermis, composed of dead skin cells and characterized by low water content. This low water content creates a gradient in the refractive index. The current work aims to elucidate the impact of a significant gradient of water content and, consequently, the variations of the refractive index of the skin on light propagation in tissues. Using analytical models of light propagation in single-layer and two-layer tissues, we predict light confinement in the stratum corneum layer. For example, the light intensity in the stratum corneum layer is noticeably (11-17%) higher than in the underlying tissue layer. This effect can be attributed to the high refractive index of the stratum corneum caused by low water content, compared with underlying tissues, and scattering in the stratum corneum layer. The effect is the most prominent for smaller diffuse reflectance of the underlying tissue. Furthermore, the effect is expected to be maximal if the thickness of the stratum corneum layer is more than the reduced scattering length. Therefore, in the visible range of the spectrum, the light confinement phenomena should be more noticeable in stratum corneum layers with a thickness of at least 150μm, which can be found in the glabrous skin of palms and soles and thickened epidermis-like calluses and corns.

1. Introduction

The epidermis is the outermost layer of the skin, and it plays a crucial role in protecting the body from external insults such as UV radiation and physical trauma. The epidermis can be subdivided into two sublayers: non-living and living epidermis. The predominant cells are the keratinocytes, arranged in five strata: the stratum corneum, stratum lucidum, stratum granulosum, stratum spinosum, and stratum basale, each with a specific function.

The topmost layer of the epidermis is called the stratum corneum, composed of dead skin cells that have migrated to the skin's surface and become flattened. This layer acts as a barrier to water loss and helps to protect the body from external insults.

Below the stratum corneum is the stratum lucidum, found only in glabrous skin. It is also composed of dead cells and packed with lipid-rich eleidin, which helps to keep water out.

The stratum corneum and stratum lucidum form the non-living epidermis. The non-living epidermis (~20 μm thick) consists of only dead squamous cells, which are highly keratinized with a high lipid (~20%) and protein (60%) content, and a relatively low (~20%) water content [ 1],

In the water content, the stratum corneum radically differs from other skin layers, which have much higher water content—the typical water content in other skin layers is 70%.

This water content gradient has interesting implications, particularly on the refractive index of the skin.

The refractive indexes of tissue structure elements, such as the fibrils, the interstitial medium, nuclei, cytoplasm, organelles, and the tissue itself, can be derived using the law of Gladstone and Dale, which states that the resulting value represents an average of the refractive indices of the components related to their volume fractions. The simplest approach will consider the tissue as a mixture of proteins and water. Assuming that proteins have a refractive index of 1.5, we can use the following expression, which is a generalized version of [2], which used water content, $c_w=0.7$

$$n_{skin} = (1-c_w)1.5 + c_w n_{water} \qquad (1)$$

The refractive index of water can be estimated as 1.333. However, there are analytical approximations that account for its dependence on the light wavelength [2].

Due to the significant difference in refractive index between proteins and water, we have a situation where tissues can be stratified into two layers: stratum corneum with the higher refractive index and all other underlying tissue with lower refractive index.

While stratum corneum has high lipid content, it does not change our considerations, as lipids also have a high refractive index.

Low water content of stratum corneum has significant implications on water content imaging [3], particularly for thickened areas (corns and calluses). However, the impact of the overall impact of low water content has not received proper attention in literature.

The current work aims to elucidate the impact of a significant gradient of water content and, consequently, the gradient of the refractive index of the skin on light propagation in tissues. To do it, we have built simplified analytical models to simulate light propagation in simplified geometries.

## 2. Methods
### 2.1. Single-Layer Model

We can consider the following model to account for the effect of mismatched boundary conditions. A semi-infinite homogeneous tissue with some optical parameters (a black box) and refractive index $n_2$ is illuminated with homogeneous wide-area illumination with light intensity $I_{01}$. Suppose we consider energy flux across a small area element $dA$ parallel to the tissue surface. In that case, hemispherical fluxes can be defined as energy fluxes across $dA$, either upward or downward. Thereafter we will refer to these hemispherical fluxes as flux, and light intensity will refer to energy transport per unit area. We also will use the following notation: the first subindex is the layer number (0 – air, 1- stratum corneum, or SC, 2- all underlying tissues); the second subindex is the light flux direction (1- downward, 2-upward). Thus, immediately above the surface, we will have incoming flux (external illumination) $I_{01}$ and outbound (reflected) flux $I_{02}$. The single-layer model is depicted schematically in Figure 1A.

Just below the surface in layer 2, we will have inbound flux $I_{21}$ and outbound (reflected) flux $I_{22}$

We can write the following equations, which connect these hemispherical fluxes

$$I_{02} = I_{01}r_{02} + I_{22}(1 - r_{20}) \qquad (2)$$
$$I_{21} = I_{01}(1 - r_{02}) + I_{22}r_{20} \qquad (3)$$
$$I_{22} = I_{21}R \qquad (4)$$

Here $r_{02}$ and $r_{20}$ are the coefficients of specular reflection from air to media and from media to air, respectively; $R$ is the diffuse reflectance of tissue calculated without a mismatched boundary. So in some sense, $R$ is the "true" diffuse reflectance, which does not consider any boundary effects.

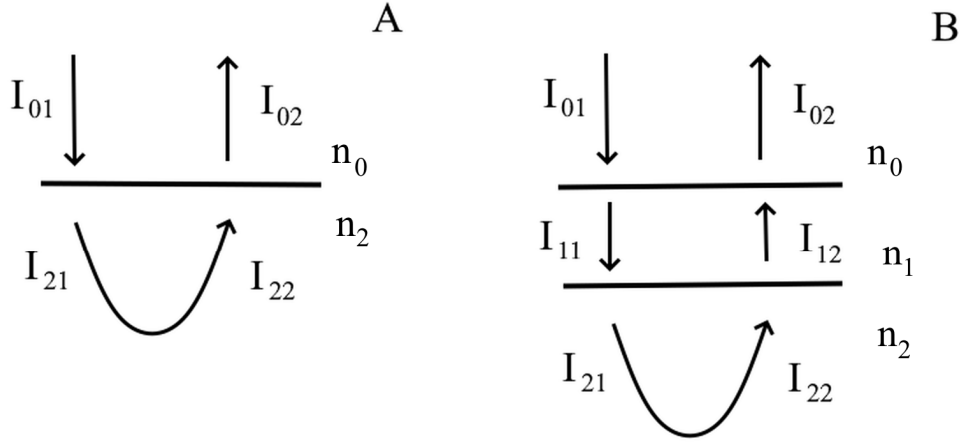

Figure 1. A schematic representation of the single-layer model (panel A) and two-layer model (panel B). The single-layer model consists of underlying tissues with a refractive index of $n_2$. The stratum corneum (SC) layer with refractive index $n_1$ covers underlying tissues in the two-layer model.

### 2.2. Two Layer Model

We can introduce a thin stratum corneum (SC) layer with a refractive index of $n_1$ between air and semi-infinite tissue to build the two-layer model on top of the single-layer model. The SC layer is typically characterized by low absorption and strong scattering. In that case, we can make an assumption that the overall light intensity within the SC layer will be homogeneous. Thus, it can be characterized by incoming flux (external illumination) $I_{11}$ and outbound (reflected) flux $I_{12}$. The two-layer model is depicted schematically in Figure 1B.

Similarly to Equations 2-4, we can write expressions for hemispherical fluxes for our two-layer model

$$I_{02} = I_{01}r_{01} + I_{12}(1 - r_{10}) \quad (5)$$
$$I_{11} = I_{01}(1 - r_{01}) + I_{12}r_{10} \quad (6)$$
$$I_{12} = I_{11}r_{12} + I_{22}(1 - r_{21}) \quad (7)$$
$$I_{21} = I_{11}(1 - r_{12)} + I_{22}r_{21} \quad (9)$$
$$I_{22} = I_{21}R \quad (9)$$

## 3. Results

### 3.1. Single layer model

Fluxes $I_{02}$, $I_{21}$, and $I_{22}$ can be expressed through external illumination $I_{01}$. However, the important values which characterize the flux distribution are the total reflectance of the tissue $r = I_{02}/I_{01}$

$$r = r_{02} + R\frac{(1 - r_{02})(1 - r_{20})}{1 - Rr_{20}} \quad (10)$$

And the ratio of the total flux just below the surface: $I_2 = I_{21} + I_{22}$ to the external light intensity:

$$I_2/I_{01} = \frac{I_{21} + I_{21}}{I_{01}} = (1 + R)\frac{1 - r_{02}}{1 - Rr_{20}} \quad (11)$$

Equation 10 is a well-known expression [4]. The implications of Equation 11 are also well-known (see, for example, [5]): the light intensity just below the mismatched surface is higher than in the case of matched boundary conditions. It can be attributed to the total internal reflectance on the boundary between more dense (tissue) and less dense (air) media.

### 3.2. Two-layer model

All fluxes in the two-layer model can be expressed through the external illumination intensity $I_{01}$. However, similarly to the single-layer model, we will pay attention to total reflectance

$$r = r_{01} + \left\{r_{12} + R\frac{(1-r_{12})(1-r_{21})}{1-Rr_{21}}\right\}\frac{(1-r_{01})(1-r_{10})}{1-r_{10}\left\{r_{12} + R\frac{(1-r_{12})(1-r_{21})}{1-Rr_{21}}\right\}} \quad (12)$$

and the ratio of total flux just below the surface ($I_1 = I_{11} + I_{12}$ in this case) to the external light intensity:

$$I_1/I_{01} = \left\{1 + r_{12} + R\frac{(1-r_{12})(1-r_{21})}{1-Rr_{21}}\right\}\frac{(1-r_{01})(1-r_{10})}{1-r_{10}\left\{r_{12} + R\frac{(1-r_{12})(1-r_{21})}{1-Rr_{21}}\right\}} \quad (13)$$

Note that Equations 12 and 13 are very similar to Equations 10 and 11. Instead of $R$, they have $r_{12} + R\frac{(1-r_{12})(1-r_{21})}{1-Rr_{21}}$. Thus, expectedly, we substitute the diffuse reflectance, $R$, in the single-layer model with the total reflectance (see Equation 10) on the interface between the stratum corneum and underlying tissues for the two-layer model.

In Figure 2, we plotted the total reflectance's dependence in single- and two-layer models as a function of "true" diffuse reflectance $R$. The refractive indexes for stratum corneum and underlying tissue were emulated using Equation 1 with a water content of 0.2 and 0.7, respectively.

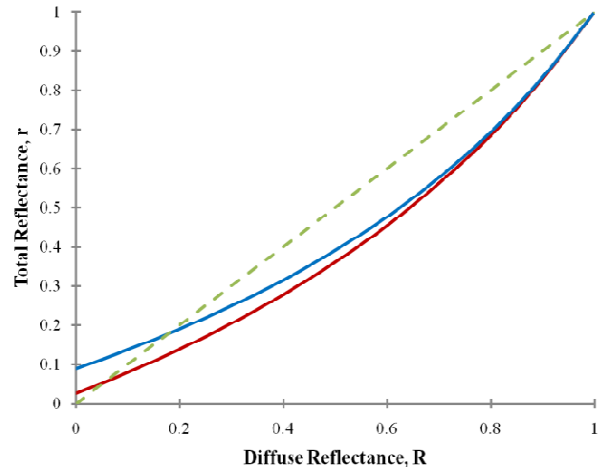

Figure 2. The total reflectance in single-layer (red line) and two-layer (blue line) models as a function of "true" diffuse reflectance $R$. The refractive indexes for stratum corneum and underlying tissue were emulated using Equation 1 with water content 0.2 and 0.7, accordingly. The diffuse reflectance, $R$, was added for visual guidance (green dashed line).

In Figure 3, we have plotted the dependence of the ratio of under-the-surface light intensity to the external light intensity in single- and two-layer models as a function of "true" diffuse reflectance $R$. The refractive indexes for stratum corneum and underlying tissue were emulated using Equation 1 with water content

0.2 and 0.7, accordingly. For comparison, we also calculated the ratio for the single-layer model if the whole tissue has low water content ($c_w$ =0.2).

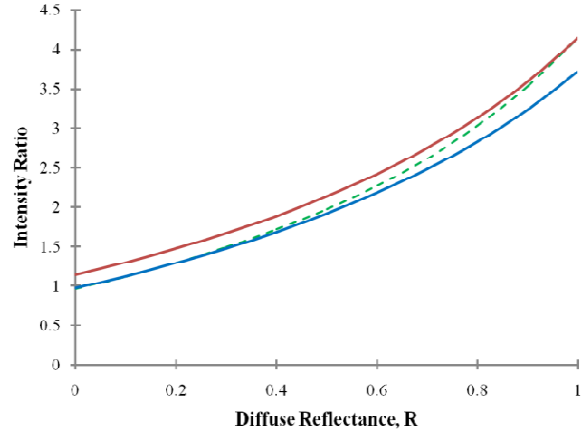

Figure 3. The ratio of under-the-surface light intensity to the external light intensity in the single-(blue line) and two-layer (red line) model as a function of "true" diffuse reflectance $R$. The refractive indexes for stratum corneum and underlying tissue were emulated using Equation 1 with water content 0.2 and 0.7, accordingly. Green dashed line: the ratio for the single layer model if the whole tissue has low water content ($c_w$=0,2)

We can also calculate the ratio of total flux just below the SC/underlying tissue interface ($I_2=I_{21}+I_{22}$) to the external light intensity:

$$I_2/I_{01} = (1 + R)\frac{1 - r_{12}}{1 - Rr_{21}}\frac{1 - r_{01}}{1 - r_{10}\left\{r_{12} + R\frac{(1 - r_{12})(1 - r_{21})}{1 - Rr_{21}}\right\}} \tag{14}$$

In Figure 4, one can see the ratio of under-the-interface light intensity in the stratum corneum $I_1$ (dashed blue line) and underlying tissue $I_2$ (red line) to the external light intensity as a function of "true" diffuse reflectance $R$ in the two-layer model calculated using Equations 13 and 14, accordingly. The light intensity in the SC layer is noticeably higher than in the nearby underlying tissue layer.

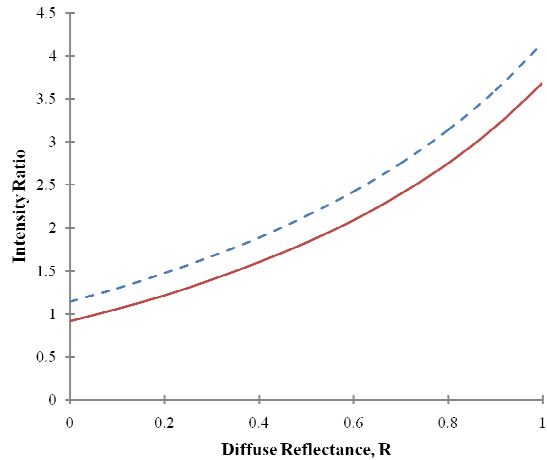

Figure 4. The ratio of under-the-interface light intensity in stratum corneum (dashed blue line) and underlying tissue (red line) to the external light intensity as a function of "true" diffuse reflectance $R$ in the two-layer model. The refractive indexes for stratum corneum and underlying tissue were emulated using Equation 1 with water content 0.2 and 0.7, accordingly.

## 4. Discussion

Our analysis shows that we can expect several noticeable effects caused by the mismatched boundary between the stratum corneum and underlying tissues.

Firstly, there is a significant increase in total reflectance for small underlying tissue diffuse reflectance values. It has interesting implications. The total reflectance of the single-layer model significantly deviates from the "true" diffuse reflectance, particularly around $R$=40-60%. Thus, any measurements of tissue reflectance need to be corrected on specular reflectance. However, for the two-layer model with stratum corneum, the total reflectance increases and is much closer to the "true" diffuse reflectance, particularly in the biologically relevant reflectance range. Thus, the necessity of correction is less obvious, which agrees with empirical observations.

Secondly, the light intensity is amplified under the surface, which can be measured by the ratio of under-the-surface light intensity to the external light intensity. In the case of matched boundaries, the ratio will be equal to *1+R*. For mismatched boundaries, the ratio typically has higher values (amplification) than predicted by *1+R*. For example, for the single-layer model, the ratio can be estimated using Equation 11.

In the proposed two-layer model, we have an additional amplification, as seen in Figure 4. The light intensity in the SC layer is noticeably higher than just below the interface with underlying tissues. This phenomenon can be attributed to several factors. The first factor is the higher value of the refractive index of the stratum corneum. However, while light intensities in the underlying tissue layers are identical in single- and two-layer models (not shown), the light intensity in the stratum corneum layer is noticeably (11-17%) higher than in the underlying tissue layer (see Figure 3). So, we can refer to it as the confinement of light in the stratum corneum layer. The higher refractive index cannot explain the full effect by itself. Figure 3 shows that the ratio is consistently higher in the two-layer model than in the one-layer model with the same (high) refractive index. The effect is the most prominent for smaller diffuse reflectance (18% increase) and disappears with its growth.

To some extent, the confinement effect is similar to fiber optics, where the core principle is in the light's confinement within the optical fiber's core. In a typical fiber optic scenario, the optical fiber consists of a core with a high refractive index, surrounded by the cladding, a material with a lower index of refraction. Thus, the light traveling in the fiber experiences multiple total internal reflections on the interfaces between the core and cladding.

Light propagation in the stratum corneum layer is quite similar. The light in the stratum corneum experiences total internal reflections on the interfaces with air and underlying tissues. The core differences are that a) instead of highly transparent media (fiber optics), we have a highly scattering media, and b) instead of a quasi-1D case (optical fiber), we have a 2D case. However, a broad range of similar phenomena should likely be observed. For example, similarly to light leakage due to fiber bending, the same phenomena can be observed on curved surfaces like heels.

As we already mentioned, the confinement effect is the most visible for low values of the diffuse reflectance *R*. It does make sense, as in this case, the underlying tissue is not a source of recycled photons for the stratum corneum layer. Thus, all recycled photons come from the total internal reflection on the interfaces between stratum corneum/air and stratum corneum/underlying tissues. For high diffuse reflectance, the underlying tissues are the primary source of recycled photons compared with relatively weak total internal reflectance on the stratum corneum/underlying tissues interface. Thus, the effects of confinement (e.g., the additional amplification) are masked by stronger mechanisms.

Thirdly, the effect depends significantly on the thickness of the stratum corneum layer. In particular, for small (compared with the distance between consecutive scatterings) thicknesses, the photons, which enter the stratum corneum from the air, pass through the SC layer without experiencing total internal reflection. The same applies to the significant part of photons entering the SC layer from the underlying tissues. All photons from the underlying tissue (other than a minute fraction of specularly reflected ones) will enter the SC layer; however, just photons with an angle of incidence $\phi > arcsin(1/n_2)$ will experience total internal reflection on the SC/air interface. All others escape the SC through the SC/air interface immediately. However, even a small fraction of photons reflected on the air/SC interface will immediately exit the SC layer on the SC/underlying tissue interface. Thus, the light confinement will be pretty minimal. In this case, photons within the tissue almost do not feel the presence of the SC layer. Thus, the ratio is close to the value provided by Equation 11, and the gap shown in Figure 4 diminishes. Moreover, in this case, an assumption in section 2.2 on light homogeneity will not hold as the light intensity within the SC layer will be very anisotropic, with a minimal component in the lateral direction.

As the SC layer thickness increases, the entering photons start experiencing scattering events. In this case, the light becomes homogenized across various directions in the SC layer. If the thickness is larger than the reduced scattering length, the photon that enters the SC layer from any direction will be homogenized in the SC layer. Thus, the share of total internal reflection in the SC layer will be maximal. As a result, the confinement will increase to its maximum value, depicted in Figure 4.

As the reduced scattering coefficient for the SC is on the scale 15-70 $cm^{-1}$ in the visible range of the spectrum [1], the light confinement phenomena should be maximal in stratum corneum layers with a thickness of at least 150μm and 600μm in blue and red ranges of the spectrum, respectively. These values are typical for the glabrous skin of palms and soles and thickened epidermis like calluses and corns.

The predicted phenomena may have implications for applications in biospectroscopy and bioimaging. In particular, it may impact the dependence of the reflected light as a function of source-detector distance.

In future work, we plan to validate our predictions using Monte Carlo simulations of light propagation in tissues.

5. Conclusions

Using the analytical model of light propagation in tissues, we predicted light confinement in the stratum corneum layer. The effect can be attributed to the high refractive index of the stratum corneum caused by low water content, compared with underlying tissues and scattering in the stratum corneum layer. Light confinement in the stratum corneum will be maximal in cases where the thickness of the stratum corneum layer is more than the reduced scattering length. In the visible range of the spectrum, the light confinement phenomena should be maximal in stratum corneum layers with a thickness of at least 150μm.